# Monte Carlo Simulation of 6MV Elekta Synergy Platform Linac photon beam using Gate/Geant4

Yahya Tayalati[a],∗ , Samir Didi[a,b] , Mustafa Zerfaoui[a,b] and Abdellilah Moussa[a]

[a] LPMR, University Mohammed First, PO. Box 717, Oujda 60000, Morocco
[b] Regional Hassan II Oncology Center, km 8, Jerada Street, P.O. Box. 2013, Oujda 60000, Morocco

ABSTRACT

The present work is devoted to develop a computational model using the Gate Monte Carlo software for the simulation of a 6MV photon beam given by Elekta Synergy Platform medical linear accelerator treatment head. The model includes the major components of the multileaf accelerator head and a homogeneous water phantom. Calculations were performed for a photon beam with several treatment fields size ranging from $5\times5$ cm$^2$ to $30\times30$ cm$^2$ at 100 cm distance from source. The simulation is successfully validated by comparison with experimental distributions measured at the Regional Hassan II Oncology Center. Good agreement between simulations and measurements was observed, with dose differences of about 1.6% and 1.8% for depth doses and lateral dose profiles, respectively. The gamma index comparisons were also performed where more than 98% of the points for all simulations passed the standard quality assurance criteria of *3mm*/*3%*.

## 1. Introduction

Since a couple of decade, Monte Carlo simulations for radiation transport were used extensively in dosimetry and medical radiation physics as an alternative to the analytical calculations. Very accurate results are obtained with these techniques thanks to the powerful distributed Grid computing resources [1-5].

In this paper, we present the simulation of the Elekta Synergy Platform medical Linear accelerator treatment head using the last release v6.2 of GATE/Geant4 Monte Carlo (MC) [6-10]. The goal is to calculate the therapeutic Dose distributions in a water phantom keeping the accuracy of results within 2%.

Gate is an advanced C++ Opensource software developed by the international OpenGATE collaboration for nuclear medicine simulation [6]. The initial focus (in 2004) was devoted to Positron Emission Tomography (PET) and Single Photon Emission Tomography (SPECT), but later on has been extended to cover the radiotherapy experiments. Built on the top of GEANT4[11-13] simulation toolkit, GATE inherits all the well validated physics models, and provide the users with a versatile integrated specific components, to handle easily a complex geometry, sources and to extract and treat elegantly the relevant information from the simulation.

All these features, participated to the growing and fast evolving of GATE use for the wide medical physics applications, GATE in its recent version plays now a key role in the design of new medical imaging devices, in the optimization of acquisition protocols and in dose calculations for radiotherapy.

The rest of the paper is organized as follows: In section 2 we will focus on the modeling of the geometry of the Elekta synergy platform accelerator, where our simulation strategy is fully described. The next section covers the results and the comparison of simulated and measured dose profiles distributions. Finally, in section 4, conclusions are drawn.

## 2. Material and Methods

### 2.1 Accelerator Geometry

Based on the vendor detailed information, we simulated the head of the linear accelerator Elekta Synergy Platform located at the regional cancer treatment center of Oujda, by use of GATE. Fig.1 illustrate the global head structure of the linear accelerator considered in this study, simulated components include:

∗ Corresponding author. Tel.: +212 672 635 779; fax: +212 536 500 609; e-mail: tayalati@cern.ch

*X-ray target*: creates Bremsstrahlung X-rays with a thin tungsten and rhenium disk approximately 0.9 mm tick. Remaining primary electrons are absorbed in a graphite absorber inside the target.

*Primary conical collimator*: made of tungsten alloy, about 10.1 cm high, located just below the x-ray target used to collimate the x-ray in the direction of the treatment field and to reduce the leakage radiations from the x-ray sources.

*X-ray beam flattening filter*: made of stainless steel, and having an accurately defined surface configuration, attached to the lower end of the primary collimator, providing uniform radiation intensity distribution across x-ray fields at any depth of treatment.

*Transmission chambers*: used for rigorous beam dose control and for the monitoring of photon and electron radiation beam output. A ceramic motherboard and a number of signal and polarising Mylar films separated by spacers made of aluminum alloy or ceramic.

*Backscatterplate*: used to avoid the excessive backscattered radiations from the secondary collimators.

*Thin Mylar Mirror*: placed on the beam central axis under the dose monitor, to enable patient set-up, and to show the position of the radiation beam.

*Multi-Leaf Collimator (MLC)*: for precision delivery of treatment and for most accurate conformal beam shaping for IMRT treatments.

*Asymmetric jaws*: used to set the overall size of the treatment field. The pair of jaws are made of tungsten about 7.8 cm tick.

**2.2 Simulation strategy**

The Gate simulation were performed in two steps: First, the accelerator patient-independent part which corresponds to the accelerator head above the secondary collimator was simulated and secondly a Phase space (Phs) was built. Gate offers the possibility to attach a phase space to any geometry in the simulation, here it is considered as a cylinder of 20 cm diameter, 1mm tick, used to store all the cinematic and production proprieties of particles above the secondary collimator. Once the first step achieved, inputs from the phase space have been used to initiate the simulation of the patient-dependent part which corresponds to photon interactions from the secondary collimator to the phantom.

Phase space build-up step is highly CPU time consuming, so that to perform this step in a reasonable time scale the corresponding task has been automatically split into sub-tasks and run on the grid computing resources located at the IN2P3 Computing Centre facility [14]. Once it is done, the subset of phase space output files were compressed and transferred to our national Moroccan grid site [15], where the next step of the simulation is proceeded using our grid computing resources. The final step consists in exploring the events using the ROOT [17] data analysis platform.

The critical parameters of the primary electron beam hitting the target, including its energy, angular distribution, and spatial distribution, were tuned using several optimization light simulation in order to minimize the simulation and measured data discrepancies. A mean electron beam energy of 6.7 MeV, 3% FWHM of the mean energy and a FWHM electron spot of 0.463 mm have been found to be in a very good agreement with the measurements.

Gate has been set up using the default parameters with an additional StepLimiter of 1 mm in water corresponding to the energy cuts of roughly 350 keV nad 5 keV for both electrons and positrons, and for photons, respectively. Furthermore, the selective Bremsstrahlung splitting [16] variance reduction technique has been also used with a splitting factor of 100 to enhances the simulation efficiency as described in [4].

In what follows, results are presented for several treatment fields size ranging from 5×5 cm$^2$ to 30×30 cm$^2$, considered with the square field defined by the X and Y Jaws at the isocenter. Comparison to real data is made using both PDDs and in-plane profiles at the following depths of 50 mm, 100 mm, 200 mm and 300 mm.

3. Results and discussion

The Gate simulation performed on grid last approximately two days. Thanks to the powerful grid tools we have been able to process a total of three billion ($3\times10^9$) primary electrons used in the work presented here. The details are summarized as follows:
Dose distributions are calculated in a 58×58×43cm$^3$ water phantom, positioned at 100cm source to surface distance (SSD). with dosel of 2 cm × 2 cm × 0.2 cm, the smallest dimension (0.2 cm) is in the measurement direction. Depth-dose curves were normalized to their maximum value at a depth Z=1.5cm. Lateral profiles were normalized to the maximum dose during the quality assurance measurements.

The Simulation efficiency has been evaluated by calculating the error estimator,

$$\varepsilon = \sum_i \frac{|D_i - Dref_i|}{Dref_i} \quad (1)$$

We have also used the gamma index [18-20] to compare the data distributions. This index takes into account both the relative shift in terms of intensity and in terms of position. For each profile, we have calculated the number of points passing the standard quality assurance criteria of 3mm/3%.

Due to the lack of primary event interactions in the tail of the penumbra of the radiation field, results are far from the measured one and are not reliable. Therefore, and to avoid a non representative increase of the statistical uncertainties, all points in this low dose area having a dose lower than 10% of the maximum dose were not considered in the comparisons.

*Field size $5\times5cm^2$*
Fig. 2 (a) shows the depth-dose profile (PDD) for both experimental and simulated data, for the 6-MV beam and SSD=100cm.
Fig. 3 (a) shows the lateral dose profiles at 50, 100, 200 and 300 mm depth, for the 6-MV beam and SSD=100cm.

*Field size $10\times10cm^2$*
Fig.2 (b) shows the depth-dose profile (PDD) for both experimental and simulated data, for the 6-MV beam and SSD=100cm.
Fig. 3 (b) shows the lateral dose profiles at 50, 100, 200 and 300 mm depth, for the 6-MV beam and SSD=100cm.

*Field size $20\times20cm^2$*
Fig. 4 (a) shows the depth-dose profile (PDD) for both experimental and simulated data, for the 6-MV beam and SSD=100cm.
Fig. 5 (a) shows the lateral dose profiles at 50, 100, 200 and 300 mm depth, for the 6-MV beam and SSD=100cm.

*Field size $30\times30cm^2$*
Fig. 4 (b) shows the depth-dose profile (PDD) for both experimental and simulated data, for the 6-MV beam and SSD=100cm.
Fig. 5 (b) shows the lateral dose profiles at 50, 100, 200 and 300 mm depth, for the 6-MV beam and SSD=100cm.

One can see that all simulated depth dose profiles are in good agreement with measurements, the uncertainty of the simulation is at whole less than 1.6%. the simulation reproduces quite well the buildup effect, Furthermore, all curves had more than 98% of points passing the gamma index comparison.

Similarly, the simulation fit accurately all measured lateral dose profiles, with no more than 1.8% of error uncertainties. Here also, all the curves fulfill the medical quality assurance criteria of *3mm/3%* with a score of 100%.

Quantitative results of the statistical comparison are summarized in table 1 and proves the simulation success.

## 4. Conclusion

This paper has presented a Gate based simulation tool that has been developed. It has been demonstrated that this tool is capable to perform an accurately simulating the dosimetric properties of a 6 MV Elketa Synergy Platform medical linac. Given the flexibility of GATE, the key components of the accelerator head have been more precisely modeled based on the manufacturer's specifications.

The results obtained with Gate simulation in terms of depth dose and lateral beam profiles for different therapeutic field sizes in a water phantom, show an excellent agreement with the measured data found with an incident mean electron energies of 6.7 MeV, and a FWHM electron spot of 0.463 mm.

This preliminary study demonstrates that GATE can be used for radiation therapy applications. Its simple macro file structure significantly facilitates the elaboration of GEANT4 simulations. Further validation will be performed with different energies and non-squared fields.

**Acknowledgments**

This project is funded by the National Center for Scientific and Technical Research CNRST (URAC07), the High Energy Physics Cluster (RUPHE) and the Spanich-Moroccans PCI2012 (A1/035250/11). The authors would like to thank the openGate collaboration for providing the toolkit and examples, regular updates, documentation, and the online user forum. The authors would like also to thank Prof. Gorge Dietmar for useful discussions and Elekta's research director Kevin Brown for the given information.

**References and notes**

1. Verhaegen F, Seuntjens V. Monte Carlo modeling of external radiotherapy photon beams.Phys Med Biol2003;48:R107–64.
2. Nederlandse Commissie Voor Stralings dosimetrie, Monte Carlo Treatment Planning: An introduction NCS Delft The Netherlands, Report 16; 2006.
3. Sardari D, Maleki R, Samavat H and Esmaeeli A. Measurement of depth-dose of linear accelerator and simulation by use of Geant4 computer code. Reports of practical oncology and radiotherapy 15 (2010)64–68.
4. Grevillot L, Frisson T, Maneval D, Zahra N, Badel J-N and Sarrut D. Simulation of a 6 MV Elekta Precise Linac photon beam using GATE/GEANT4. Phys. Med. Biol.56(2011) 903–918.
5. Cornelius I, Hill B, Middlebrook N, Poole C, Oborn B and Langton C. Commissioning of a Geant4 based treatment plan simulation tool: linac model and dicom-rt interface. http://eprints.qut.edu.au/41556/.
6. Jan S et al., GATE V6: a major enhancement of the GATE simulation platform enabling modelling of CT and radiotherapy. Phys. Med. Biol. 56 (2011) 881-901.
7. Jan S et al., GATE: a simulation toolkit for PET and SPECT. Phys. Med. Biol. 49 (2004) 4543-4561.
8. Strul D, Santin G, Lazaro D, Breton V, Morel C. GATE (Geant4 Application for Tomographic Emission): a PET/SPECT general-purpose simulation platform. Nucl. Phys. B (Proc. Suppl.) 125 (2003) 75-79.
9. Santin G, Strul D, Lazaro D, Simon L, Krieguer M, Vieira M, Breton V, Morel C. GATE: A Geant4-based simulation platform for PET, SPECT integrating movement and time management. IEEE Trans. Nucl. Sci. 50 (2003) 1516-1521.
10. The Gate web site, http://www.opengatecollaboration.org/.
11. Allison J et al., Geant4 Developments and Applications, IEEE Transactions on Nuclear Science 53 No. 1 (2006) 270-278.
12. Agostinelli S et al., Geant4 - A Simulation Toolkit, Nuclear Instruments and Methods A 506 (2003) 250-303.
13. The Geant4 web site http://geant4.cern.ch/.
14. IN2P3 Computing Centre facility web site http://cc.in2p3.fr/.
15. The national Moroccan grid center http://www.magrid.ma/.
16. Kawrakow I, Rogers D W and Walters B R 2004 Large efficiency improvements in BEAMnrc using directional bremsstrahlung splittingMed. Phys.312883–98.
17. Brun R and Rademakers F, ROOT - An Object Oriented Data Analysis Framework, Proceedings AIHENP'96 Workshop, Lausanne, Sep. 1996, Nucl. Inst. & Meth. in Phys. Res. A 389 (1997) 81-86. See also http://root.cern.ch/.
18. Daniel A. Low and James F. Dempsey, Evaluation of the gamma dose distribution comparison method, Medical Physics, Vol. 30, No. 9, September 2003.
19. Daniel A. Low, William B. Harms, Sasa Mutic, and James A. Purdy, A technique for the quantitative evaluation of dose distributions, Medical Physics, Vol. 25, No. 5, May 1998.
20. Bak J, Choi JH, Kim JS and Park SW , Modified dose difference method for comparing dose distributions, Journal of Applied Clinical Medical Physics , Vol. 13 , No. 2 , Spring 2012 .

| Field size (cm²) | Depth (cm) | Lateral dose | | Depth-dose | |
|---|---|---|---|---|---|
| | | ε (%) | γ (%) | ε (%) | γ (%) |
| 5×5 | 5 | **1.7** | **100** | **1.6** | **99** |
| | 10 | **0.2** | | | |
| | 20 | **0.2** | | | |
| | 30 | **0.1** | | | |
| 10×10 | 5 | **0.5** | **100** | **0.6** | **99** |
| | 10 | **0.6** | | | |
| | 20 | **0.8** | | | |
| | 30 | **0.5** | | | |
| 20×20 | 5 | **1.2** | **100** | **0.3** | **99** |
| | 10 | **0.9** | | | |
| | 20 | **0.8** | | | |
| | 30 | **0.3** | | | |
| 30×30 | 5 | **1.8** | **100** | **0.1** | **98** |
| | 10 | **1.5** | | | |
| | 20 | **0.8** | | | |
| | 30 | **1.3** | | | |

**Table 1.** The efficiency of the simulation is presented in term of the error estimator $\varepsilon$ of eq.1 and in term of the percentage of points passing the 3%/3mm gamma comparison, for both the Lateral and depth dose curves.

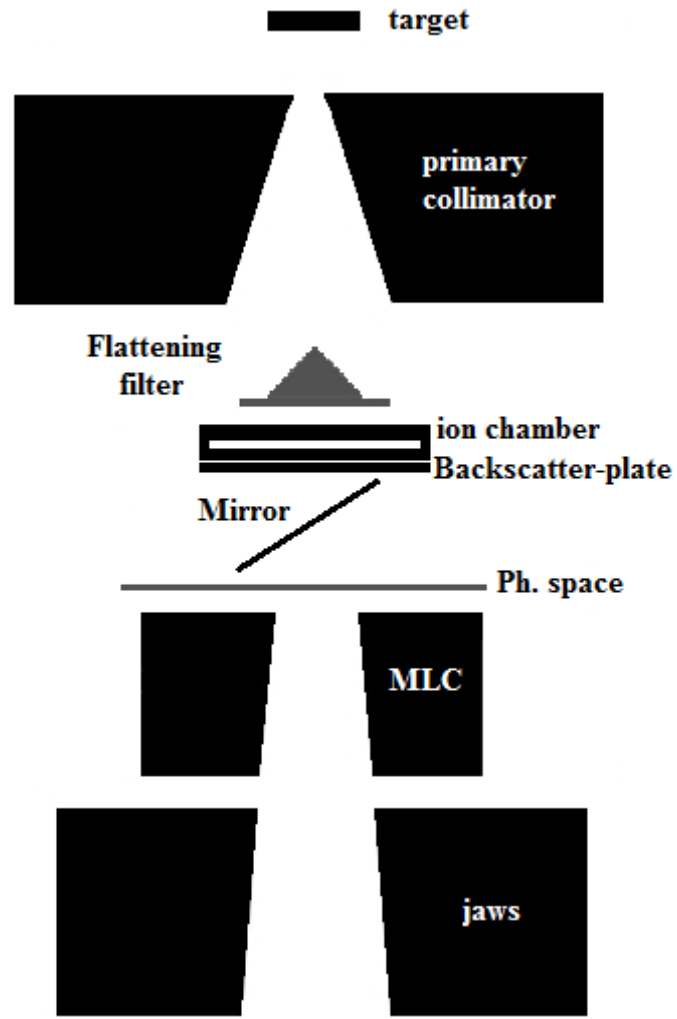

**Figure 1.** Schematic of geometry and component modules used in Gate for modeling the Elekta linac gantry for 6 MV photon beam.. The components are not to scale.

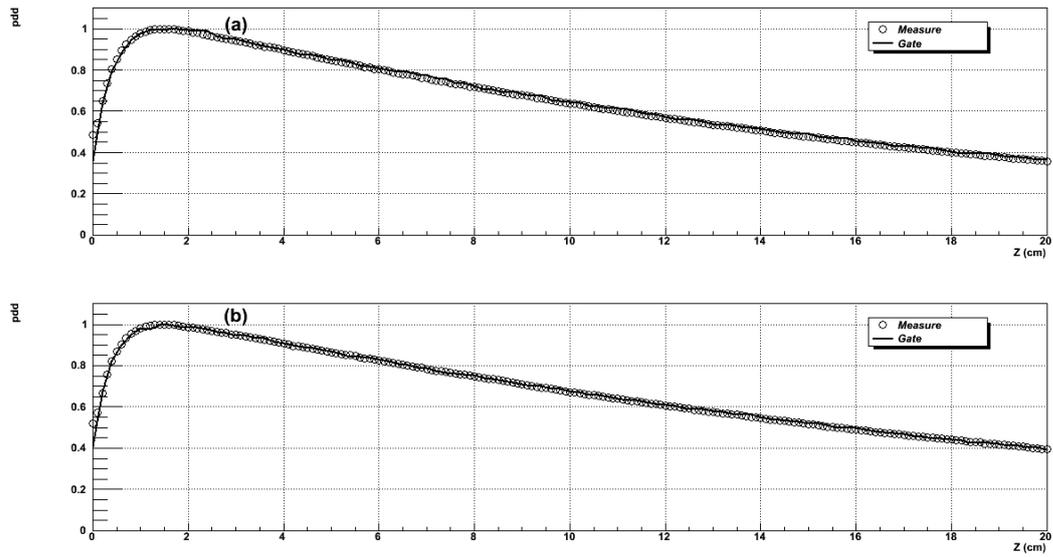

**Figure 2.** for the 5×5 cm$^2$ (a) and 10×10 cm$^2$ (b) fields, are represented the Depth-dose profile with the 6-MV beam. Circles refer to measured data and the solid lines refer to Gate/Geant4 Monte Carlo results.

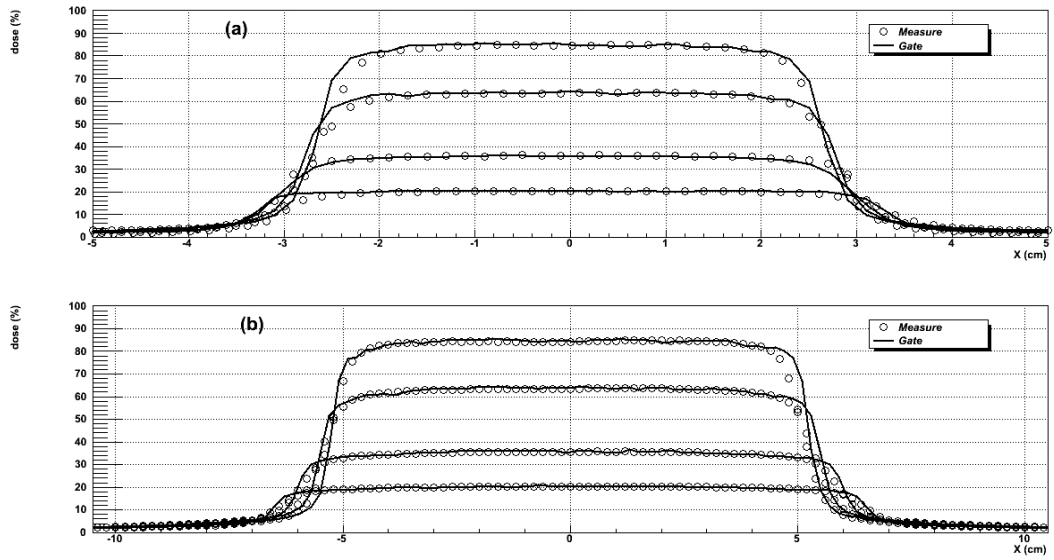

**Figure 3**. Lateral dose profiles at 50,100, 200 mm and 300 mm depth for the 5×5 cm$^2$ (a) and 10×10 (b) cm$^2$ fields with a 6-MV beam. circles refer to measured data, and the solid lines refer to GATE/Geant4 Monte Carlo results.

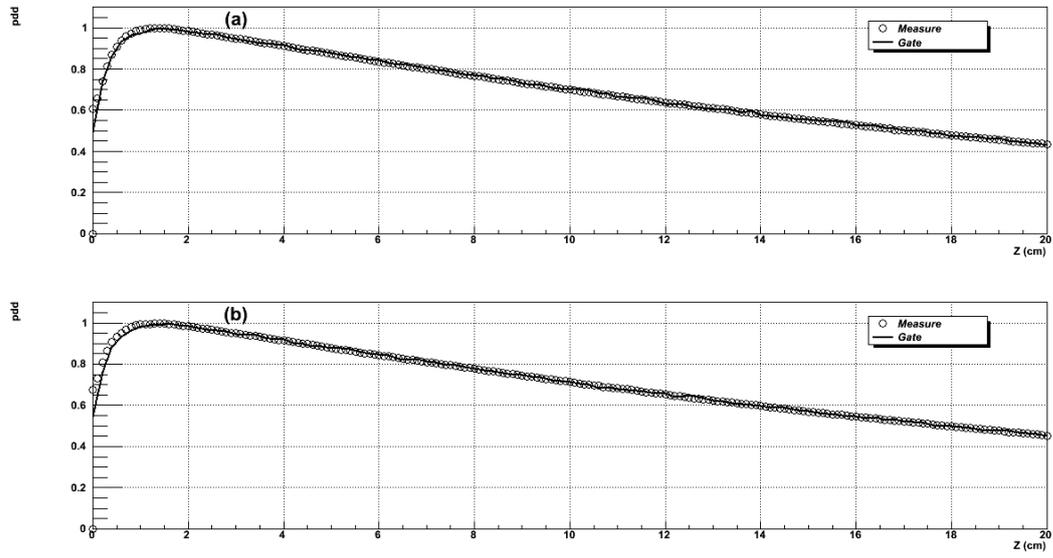

**Figure 4.** Same as figure 2 for the 20×20 cm$^2$ (a), and the 30×30 cm$^2$ (b) fields.

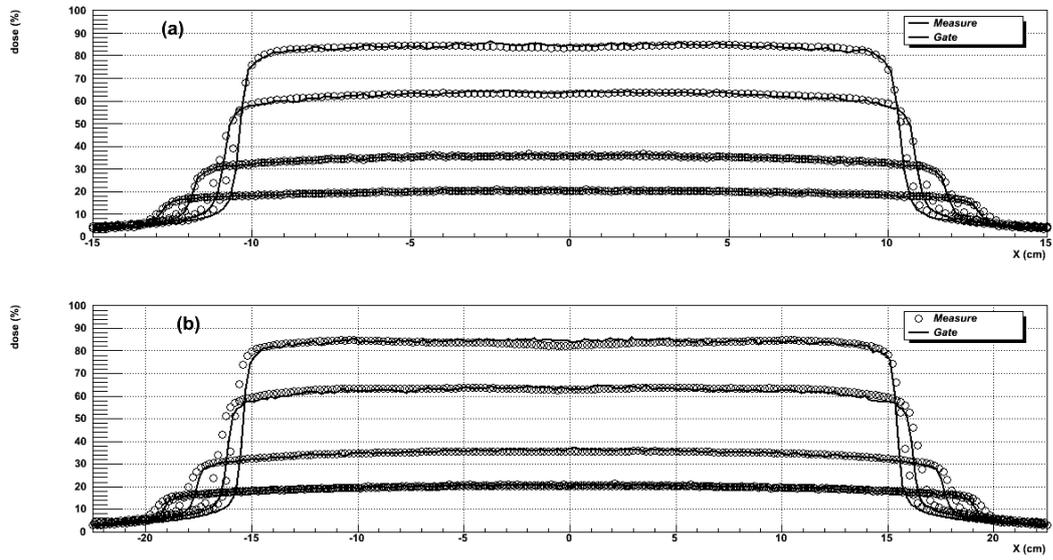

**Figure 5.** Same as figure 3 for the 20×20 cm² (a), and the 30×30 cm² (b) fields.